\documentclass[aps,pra,reprint,amsmath,amssymb,floatfix]{revtex4-2}

\usepackage{graphicx,dblfloatfix}% Include figure files
\usepackage{dcolumn}% Align table columns on decimal point
\usepackage{bm}% bold math

\usepackage[utf8]{inputenc}
\usepackage{enumitem}
\usepackage[T1]{fontenc}
\usepackage{mathptmx,etoolbox,comment,amsmath,siunitx}
\usepackage{booktabs}
\usepackage[hidelinks]{hyperref}

\usepackage[
    starfontserif % comment for sans glyphs
    ]{starfont}
\DeclareSymbolFont{starfontsym}{OT1}{sts}{m}{n}
\DeclareMathSymbol{\mathTerra}{\mathord}{starfontsym}{76}
\DeclareMathSymbol{\mathvarTerra}{\mathord}{starfontsym}{108}

\makeatletter
\def\@email#1#2{%
 \endgroup
 \patchcmd{\titleblock@produce}
  {\frontmatter@RRAPformat}
  {\frontmatter@RRAPformat{\produce@RRAP{*#1\href{mailto:#2}{#2}}}\frontmatter@RRAPformat}
  {}{}
}%
\makeatother
\begin{document}

\title[]{Quantifying structural nonlinearity in a disordered Fabry-Pérot cavity 
%Quantifying structural nonlinearity: theory and experiment in a disordered Fabry-Pérot cavity
} 
% Force line breaks with \\ 
\author{Morgan Facchin$^*$} 
\email{Contact author: morgan.facchin@lkb.ens.fr} 
\author{Sylvain Gigan} 
\affiliation{Laboratoire Kastler Brossel, École Normale Supérieure - Université Paris Sciences et Lettres, Sorbonne Université, Collège de France, CNRS, UMR 8552, Paris 75005, France} 
%\affiliation{-----} 
%\date{\today} 

\begin{abstract} 
Structural nonlinearity is the emergence of nonlinear input-output transformations from purely linear processes. In optics, this arises from the multiple interactions of light with input data. Current implementations achieve either a finite number of interactions, or partial modulation of the optical field. In this study, we present an architecture that combines both arbitrarily many interactions and full modulation. Our design consists of a Fabry-Pérot cavity, where one mirror is replaced by a spatial light modulator on which data is displayed. We develop an analytical model from which we derive closed-form expressions for the main metrics of nonlinearity, and find good agreement with experiment. Our analysis extends beyond the present implementation and reveals relationships that are independent of a specific physical system, providing a basis for a unified description of structural nonlinearity, and a first-principles design guide for machine learning applications.

\end{abstract}
\maketitle

\section{Introduction}
Recent years have seen growing interest in optical computing, which aims to perform computations using light rather than conventional electronics, for its potential advantages in speed and energy efficiency \cite{mcmahon2023physics,wetzstein2020inference,farmakidis2024integrated}. These advantages have been demonstrated particularly for linear transformations, including random projections \cite{saade2016random,ohana2023linear}, matrix–vector multiplication \cite{zhou2022photonic}, convolution \cite{feldmann2021parallel,xu2021tops}, and various linear components of optical machine learning \cite{shen2017deep,larger2017high,lin2018all,hamerly2019large,rafayelyan2020large,dong2025high,wang2026streamlined}. However, complex computational tasks fundamentally rely on nonlinear transformations. These have been implemented through various nonlinear optical effects \cite{zuo2019all,wang2024large}, but energy efficiency is a common limitation in such implementations, which hinders the initial motivation for optical computing. A common alternative has been to use cascaded optics-electronics-optics conversions, adding significant energy overhead and latency. 

A promising solution was recently proposed: structural nonlinearity. Structural nonlinearity is an approach for implementing nonlinear transformations using only linear processes \cite{eliezer2023tunable,xia2024nonlinear,yildirim2024nonlinear,wanjura2024fully}, thus providing significantly more expressivity than a single-layer network, without requiring either complex optical non-linearity or optics-electronics-optics conversions. Data is encoded into the structure of an otherwise linear system (using for example a spatial light modulator), and the nonlinearity arises from the multiple interactions of light with the data. Whenever light interacts with the data more than once, products of the data appear in the optical field, producing higher-order terms.
The effect has been explored experimentally in an integrating sphere \cite{eliezer2023tunable,xia2024nonlinear}, multiple-bounce geometries \cite{yildirim2024nonlinear,xia2025reconfigurable,venancio2026optical}, a thin LCD cavity \cite{liu2026nonlinear}, and fiber-loop optical networks \cite{wu2026time}. 
Numerical studies include integrated photonics \cite{wanjura2024fully}, diffractive networks \cite{li2024nonlinear,wu2025coupling}, optical Kolmogorov-Arnold Networks \cite{stroev2026programmable}, programmable metasurfaces \cite{hammami2026expressivity}, alongside theoretical investigations of universality~\cite{savinson2025universality}, and training \cite{wanjura2024fully}.

For maximizing nonlinearity, one seeks to maximize: 1. the amount of modulation applied to the optical field, and 2. the number of interactions of light with the input data. In multiple-bounce geometries, where light reflects several times on light-modulating devices (SLMs or DMDs), the first condition is met, but the number of interactions with the data remains finite (with a maximum of four, to our knowledge, in \cite{yildirim2024nonlinear}). An unbounded number of interactions is desirable for expressivity and, potentially, reaching universality \cite{savinson2025universality}. Multiple-scattering geometries in complex media (such as the integrating sphere) overcome this finite-order limitation by allowing light to revisit the modulated region an arbitrarily large number of times, although with exponentially vanishing amplitude. However, these only allow a small fraction of the system to be modulated (a few percent). Moreover, multiple-scattering geometries are most prone to perturbations \cite{facchin2024determining}, in particular thermal expansion. In integrating spheres, the optical power of the input laser alone is sufficient to decorrelate the speckle within tens of minutes \cite{Facchin2021ref}.

In this study, we introduce a new architecture that provides both full modulation and an unbounded number of interactions. Our design consists of a Fabry-Pérot cavity, where one of the mirrors is replaced by a Spatial Light Modulator (SLM) on which input data is displayed, hence coined \textit{disordered} Fabry-Pérot cavity. Multiple reflections allow light to interact with the data an arbitrarily large number of times, while the SLM makes the cavity an entirely modulated system, with no uncontrolled fixed parts.

\section{Experimental setup  }
\begin{figure}[h!] 
\centering\includegraphics[width=1\columnwidth]{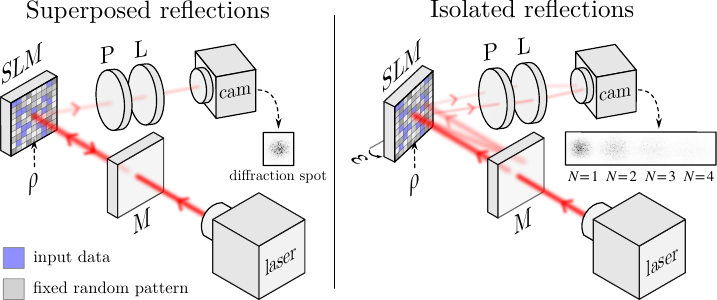 }
\caption{\textbf{Experimental setup.} Laser light enters a cavity formed by a planar mirror (M) and an SLM of reflectivity $\rho$, where it undergoes multiple reflections. A single diffraction order of the SLM is selected, passed through a polarizer (P) and a lens (L) and focused onto a camera. Two configurations are possible. \textit{Superposed reflections}: The mirror and SLM are parallel, causing all diffracted light to superpose into a single spot. \textit{Isolated reflections}: The SLM is tilted by a small angle $\epsilon$, causing each reflection to diffract at an increasingly larger angle, resulting in equally spaced diffraction spots. Input data is displayed on a fraction $\alpha$ of the SLM, on top of a fixed random phase pattern that ensures the randomization of light.}
\label{setup}
\end{figure}

The optical setup is described in Fig.~\ref{setup}. The source is a continuous-wave 532~nm laser (Oxxius LAS-01599) with a power of 50~mW and a \(4\sigma\) beam diameter of 3~mm. Light enters the 4-cm-long cavity through the rear side of a planar dielectric mirror. 
To minimize intracavity losses, the mirror is chosen to be highly reflective, with a specified reflectivity of 0.996, so that only 0.4\% of the incident power (about 0.2~mW) enters the cavity and undergoes multiple reflections. To prevent back-reflections into the laser, the mirror is slightly tilted so that the reflected beam is intercepted by a beam block. The SLM is a Holoeye Pluto 2 with a resolution of \(1920\times1080\) pixels and a pitch of \qty{8}{\micro\metre}. A linear polarizer placed before the cavity (not shown) ensures pure phase modulation on the SLM. Phase patterns spanning a \(2\pi\) range are displayed using \(12\times12\)-pixel macropixels.\\

A delicate aspect of such a system is extracting light from the cavity without significantly increasing loss. This is achieved here by exploiting the intrinsic diffraction produced by the SLM: After each reflection, a fraction of light is inevitably lost into the diffraction orders, which is repurposed here as our probe light. A single diffraction order (approximately the 10th, which has a large enough angle to escape the cavity without touching the mirror) is selected, passed through a polarizer to remove the residual unmodulated light (which, although small, would otherwise form an intense central spot at the focal plane), and focused onto a camera (Allied Vision Mako U-051B) using a lens of 5~cm focal length. Two configurations are possible: 

\noindent\textbf{Superposed reflections:} the SLM and mirror are parallel. After each reflection, the diffracted light emerges in the same direction and is focused by the lens onto a single region of the camera, forming a single diffraction spot. In this configuration the two main experimental parameters are the fraction of modulated area ($\alpha$), and the power transmission of the cavity after one round-trip, which is essentially the reflectivity of the SLM ($\rho$), as absorption by the mirror is negligible. 

\noindent\textbf{Isolated reflections:} the SLM is tilted by a small angle~$\epsilon$. After each reflection, the diffracted light emerges at an increasingly larger angle: a beam that would be diffracted at an angle $\theta_0$ without tilt is diffracted at an angle $\theta_0+2N\epsilon$ after the $N$th reflection. As the lens maps angles to positions, we observe equidistant diffraction spots on the camera, each successive one corresponding to one additional reflection on the SLM. Note that  on the SLM plane, on the other hand, the successive reflections are not equidistant. It can be shown that the $N$th reflection is at a distance $2N(N-1)\epsilon d$ from the spot of the first incident beam ($N=1$), with $d$ the mirror-SLM distance. This sets a limit to the maximal number of reflections that can be observed, as the beam eventually walks off the SLM. In this configuration the main experimental parameters are the fraction of modulated area ($\alpha$), and the selected diffraction spot ($N$). \\

The device operates as follows. A macropixel size is first chosen to adjust the size of the diffraction spots, here \(12\times12\) pixels, for diffraction spots of 10 mrad. A fixed random phase pattern is then applied to the SLM, with phase values uniformly distributed over a \(2\pi\) interval. The input data is scaled to a \(\pi\) interval and then added, macropixel by macropixel, to this fixed pattern on a fraction $\alpha$ of the macropixels, randomly distributed over the SLM surface. The fixed pattern ensures spatial mixing of the field after each round trip, as well as fully-developed speckle statistics in each diffraction spot, independently of the input data. 

Although the isolated reflections configuration does not provide the unbounded number of interactions discussed above, it offers a unique feature that may be of interest in certain applications, which is to simultaneously capture different interaction orders in a single image. Therefore, in the following, we treat both configurations in parallel.

\section{Interaction-order spectrum}

We begin by introducing a quantity that naturally emerges in our description of the system: we define \(P_n\) as the fraction of output optical power that has interacted \(n\) times with the input data. Since each interaction applies a data-dependent factor to the field, \(P_n\) may equivalently be interpreted as the fraction of the output field that is of order \(n\) (where a term is said to be of order \(n\) if it contains a product of \(n\) data elements). In analogy with a frequency spectrum, giving the amount of each frequency in a signal, we shall refer to $P_n$ as the interaction-order spectrum, or simply \textit{order spectrum}, giving the amount of each order in the field. The usefulness of $P_n$ lies in the fact that the main metrics of nonlinearity can be derived from it, and it can be determined from simple statistical arguments, as we now show for the two configurations.  \\

\noindent In the isolated reflections configuration, when light reflects from the SLM with a fraction of modulated area $\alpha$, interaction with the data can be viewed as a Bernoulli process with success probability $\alpha$. For the $N$th reflection, this Bernoulli process is repeated $N$ times, and the probability of $n$ interactions after $N$ trials is given by the binomial law, hence  
\begin{equation} \label{binom}
P_n=\binom{N}{n}\alpha^{n}(1-\alpha)^{N-n}. 
\end{equation} 

\noindent In the superposed reflections configuration, each reflection contributes to $P_n$ according to Eq.~\eqref{binom}, but is weighted by its power, proportional to $\rho^N$. Summing over $N$ and normalizing yields
\begin{equation} \label{PN_superp}
\begin{split}
P_n=\frac{1-\rho}{\rho}\bigg(\frac{(\rho \alpha)^n}{(1-\rho(1-\alpha))^{n+1}}-\delta[n]\bigg),
\end{split}
\end{equation}
with $\delta$ the Kronecker delta symbol.

\section{An analytical model of nonlinearity}
A few metrics of nonlinearity pertinent to structural nonlinearity have been identified, in particular in \cite{han2026optical}. We consider two of these metrics, the coefficient of determination and the kernel profile, and show that they can be analytically expressed in terms of the order spectrum $P_n$. We then introduce a third metric, the average number of interactions with the data, which provides a complementary interpretation and takes a particularly simple form. \\

\noindent\textbf{The coefficient of determination $R^2$}

\noindent The coefficient of determination, or $R^2$, is a common metric of goodness of fit in regression, which can be used as a metric of nonlinearity. Indeed, a good linear fit ($R^2\approx 1$) reveals a linear input-output relationship, while lower values reveal the presence of nonlinearity. It is expressed as 
\begin{equation} \label{R2}
R^2=1-\frac{\text{Var}(\boldsymbol{u}-W\boldsymbol{x})}{\text{Var}(\boldsymbol{u})}, 
\end{equation}
where $\boldsymbol{u}$ is the output field, $W\boldsymbol{x}$ is the best linear model with weight matrix $W$, and input $\boldsymbol{x}$ (we assume phase encoding with $\boldsymbol{x}=e^{i\mathbf{\phi}}$). For a multivariate output, the variance is expressed as $\mathrm{Var}(\boldsymbol{u})=\left\langle\ |\boldsymbol{u}-\langle \boldsymbol{u}\rangle\ |^2\right\rangle$, with $|\cdot|^2$ the squared vector norm, and $\left\langle \cdot \right\rangle$ averaging over realizations of $x$. 

Note that a precaution must be taken in the empirical determination of $R^2$: if the number of samples is equal to the input dimension, $R^2$ equals 1, regardless of the input-output relationship, as the linear regression contains more parameters than unknowns. Estimating \(R^2\) therefore requires a large number of samples compared to the input dimension. This is only asymptotic, however, and Eq. (\ref{R2}) is always a biased overestimate. This bias can be corrected efficiently using the \textit{adjusted}~$R^2$ \cite{raju1997methodology}. 

By expressing the output field as a sum over all possible propagation paths and grouping terms by their number of interactions with the data, it can be shown (see Appendix A) that Eq. (\ref{R2}) takes the form  
\begin{equation} \label{R2P01}
R^2=\frac{P_1}{1-P_0}.
\end{equation}
This yields \(R^2=1\) for any order spectrum satisfying \(P_0+P_1=1\), which is consistent with a regression model containing both a constant and a linear term (strictly speaking, an affine regression). \\

\noindent For superposed reflections, inserting Eq.~(\ref{PN_superp}) into Eq.~(\ref{R2P01}) yields the closed-form expression
\begin{equation} 
R^2=\frac{1-\rho }{1-\rho(1-\alpha)  }. 
\end{equation}

\noindent Similarly for isolated reflections, inserting Eq. (\ref{binom}) into Eq. (\ref{R2P01}) yields the closed-form expression for the $N$th reflection 
\begin{equation} 
R^2=\frac{\alpha  N (1-\alpha )^{N-1}}{1-(1-\alpha )^N}. 
\end{equation}

\noindent\textbf{The kernel profile $k(C)$}

\noindent Another characterization of nonlinearity identified in \cite{han2026optical} is the \textit{kernel profile}, defined as follows. Consider a pair of input data with correlation $C$, and the corresponding pair of output fields with correlation $C'$. The kernel profile is the relationship between the two correlations, or more precisely $k(C)=\langle C'\rangle$, where the brackets denote expectation over realizations of input pairs having correlation $C$. 
The kernel profile characterizes the sensitivity of the output to changes in the input, and describes how correlations (or, equivalently, distances) are transformed. A linear transformation produces a straight kernel profile, whereas curvature indicates nonlinearity. 

Similarly to $R^2$, the kernel profile can be expressed in terms of $P_n$. It can be shown (see Appendix B) that the kernel profile of our system is given by 
\begin{equation} \label{gamma}
k(C)=\sum_{n=0}^\infty P_n \, C^n.
\end{equation}
This is a power series of the input correlation, where the coefficients are the order spectrum. The assumptions underlying this result are the following: input data is uniformly illuminated, each reflection is normal to the SLM, and full-mixing conditions are fulfilled (i.e. each pixel of the data illuminates every other pixel of the data after one round-trip). \\

\noindent For superposed reflections, inserting Eq. (\ref{PN_superp}) into Eq. (\ref{gamma}) yields the closed-form expression 
\begin{equation} \label{kernelsuperp}
k(C)=\frac{(1-\rho)(\alpha C +1-\alpha)}{1-\rho (\alpha C +1-\alpha)},
\end{equation}

\noindent Similarly for isolated reflections, inserting Eq. (\ref{binom}) into Eq. (\ref{gamma}) yields the closed-form expression 
\begin{equation} \label{kernelisol}
k(C)=(\alpha C +1 -\alpha)^N, 
\end{equation}

Note the appearance of the quantity $\alpha C +1-\alpha$ in both Eq.~\eqref{kernelsuperp} and Eq.~\eqref{kernelisol}. This quantity corresponds to the correlation of the entire surface of the SLM, while $C$ is the correlation of the modulated part only. Therefore, the kernel profiles may be expressed more simply as $(1-\rho)C_{tot}/(1-\rho C_{tot})$ and $C_{tot}^N$, respectively, with $C_{tot}$ the correlation of the entire SLM. Then the substitution $C_{tot}\xrightarrow{}\alpha C +1-\alpha$ can be interpreted as a rescaling that expresses the result in terms of the modulated part only. \\

\noindent\textbf{The average number of interactions with the data $\langle n\rangle$}

\noindent A closer analysis of the kernel profile suggests an additional metric. 
It can be noted from Eq.~\eqref{gamma} that $k(C)$ is monotonic between 0 and 1. Indeed, its derivative is
\begin{equation}
\frac{\text{d}k(C)}{\text{d}C}=\sum_{n=0}^\infty n \, P_n \, C^{n-1}, 
\end{equation}
which is always positive on that interval. The same can be said of its second derivative, from which it follows that its graph is always concave. In fact, all derivatives are positive, and $k(C)$ is a strongly constrained function. One parameter that is not constrained is its slope at $C=1$. Therefore, it can be anticipated that the general shape of $k(C)$ is strongly determined by this slope. Evaluating it at $C=1$ we have: 
\begin{equation} \label{nav}
\frac{\text{d}k(C)}{\text{d}C}\Big |_{C=1}
=\sum_{n=0}^\infty n \, P_n 
\end{equation}
This quantity has a direct interpretation: it is the average number of interactions of light with the data. This is our third metric, denoted by $\langle n\rangle$. \\

\noindent For superposed reflections, we differentiate Eq.~(\ref{kernelsuperp}) and find   
\begin{equation} \label{avNsuperp}
\langle n\rangle=\frac{\alpha}{1-\rho}. 
\end{equation}

\noindent Similarly for isolated reflections, we differentiate Eq. (\ref{kernelisol}) and find
\begin{equation} \label{avNisol}
\langle n\rangle=\alpha N. 
\end{equation} 

Note that this equality between the slope of the kernel profile and $\langle n\rangle$ requires the assumptions of our model (i.e. normal incidence and full mixing). Otherwise, those are two different quantities. 
\\

\noindent\textbf{A place for each metric}

\noindent $R^2$ and $\langle n\rangle$ capture complementary aspects of nonlinearity. $R^2$ strictly quantifies nonlinearity, while $\langle n\rangle$ does not but provides a better measure of its strength in certain cases. To illustrate this, consider the two cases $\{P_0=1/2, P_2=1/2\}$, and $\{P_1=1\}$. In both cases, $\langle n\rangle$ is 1, which fails to distinguish the nonlinear nature of the former case from the fully linear nature of the latter, while $R^2$ is respectively 0 and 1. On the other hand, consider the two cases $\{P_2=1\}$ and $\{P_{10}=1\}$, both yield an $R^2$ of zero, although the latter is much more nonlinear, which is captured by $\langle n\rangle$, respectively equal to 2 and 10.

\begin{figure}[h!] 
\centering\includegraphics[width=1\columnwidth]{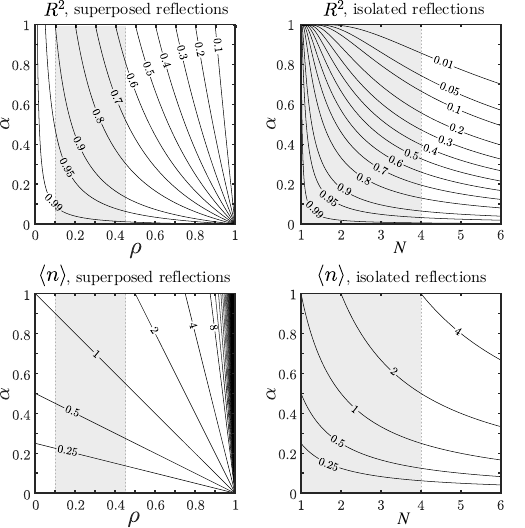 }
\caption{\textbf{Maps of $R^2$ and $\langle n\rangle$ in parameter space.}
The values of $R^2$ (coefficient of determination) and $\langle n\rangle$ (average number of interactions with the data) are plotted as isolines in $\alpha-\rho$ space for superposed reflections, and in $\alpha- N$ space for isolated reflections (where $N$ is treated as a continuous variable). The regions of parameter space accessible to our setup are shown shaded. 
}
\label{isolines}
\end{figure}

\setlength{\tabcolsep}{7pt}
\begin{table}[t]
    \label{tab:nonlinearity_metrics}
\begin{tabular}{@{}lcc@{}}
\toprule
& Superposed reflections & Isolated reflections \\
\midrule
$P_n$ &
$\displaystyle \frac{1-\rho}{\rho}\left[\frac{(\rho\alpha)^n}{\bigl(1-\rho(1-\alpha)\bigr)^{n+1}}-\delta[n]\right]$ &
$\displaystyle \binom{N}{n}\alpha^n(1-\alpha)^{N-n}$ \\[5ex]

$R^2$ &
$\displaystyle \frac{1-\rho}{1-\rho(1-\alpha)}$ &
$\displaystyle \frac{\alpha N(1-\alpha)^{N-1}}{1-(1-\alpha)^N}$ \\[4ex]

$k(C)$ &
$\displaystyle \frac{(1-\rho)(\alpha C+1-\alpha)}{1-\rho(\alpha C+1-\alpha)}$ &
$\displaystyle (\alpha C+1-\alpha)^N$ \\[5ex]

$\langle n\rangle$ &
$\displaystyle \frac{\alpha}{1-\rho}$ &
$\displaystyle \alpha N$ \\
\bottomrule
\end{tabular}
    \centering
    \caption{Summary of the analytical expressions of $P_n$ (the order spectrum), $R^2$ (the coefficient of determination), $k(C)$ (the kernel profile), and $\langle n \rangle$ (the average number of interactions), for isolated and superposed reflections.}
\end{table}

\section{Experimental verification}
\begin{figure}[h!] 
\centering\includegraphics[width=1\columnwidth]{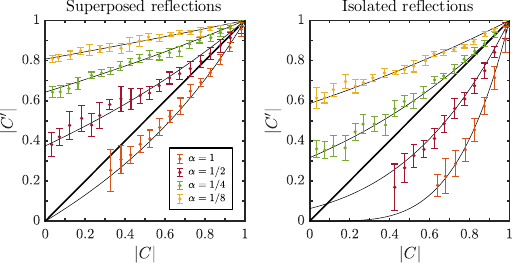 }
\caption{\textbf{Experimental kernel profiles.}
Measurements of the output correlation norm $|C'|$ as a function of the input correlation norm $|C|$ for isolated reflections (\(N=4\)), and superposed reflections (\(\rho=0.40\pm0.02\)). In each case, measurements are shown for different values of \(\alpha\) (legend applies to both panels). The identity function $k(C)=C$ (thick line) is shown for reference, together with the predictions of Eqs.~\eqref{kernelsuperp} and \eqref{kernelisol} (thin lines). More measurements are shown in Appendix D. }
\label{fig:kernels}
\end{figure}

We experimentally verify our model by measuring kernel profiles with different parameters and comparing against the predictions of Eqs.~\eqref{kernelsuperp} and ~\eqref{kernelisol}. We display a series of 30 patterns on the SLM, with controlled correlations between them, which provides a total of $30(30-1)/2=435$ distinct pairs. For each pair, we compute the input correlation $C$ between the two corresponding encoded fields, and the correlation $C_I$ between the two corresponding speckle patterns (in a 40$\times$40-pixel crop centered at the diffraction spot). For both quantities, correlation is defined as the normalized covariance (or Pearson correlation) expressed between two variables $x$ and $y$ as $C=\langle (x^*-\langle x^* \rangle)(y-\langle y\rangle) \rangle/(\sigma_x\sigma_y)$, where $\left\langle \cdot \right\rangle$ denotes spatial averaging, and $\sigma$ the standard deviation. Assuming a circular Gaussian output field (ensured by the fixed random pattern), the intensity correlation is equal to the absolute square of the field correlation $C'$ \cite[sec~3.3.4]{Goodman}. We therefore estimate the output field correlation as $|C'|=\sqrt{C_I}$. The 435 values of \(|C|\) are split into intervals of width 0.05, and the mean \(|C'|\) in each interval is plotted with its standard deviation shown as an error bar.
The resulting kernel profiles are shown in Fig. \ref{fig:kernels}, and find good agreement with the theoretical predictions. 

Controlled values of input correlation are obtained with phase patterns of the form $\phi=\text{mod}(\phi_0+a\Delta \phi,2\pi)$, with $\phi_0$ a random pattern (fixed across all 30 inputs), $\Delta \phi$ a normal Gaussian random phase pattern (different across all 30 inputs), and $a$ a constant ranging uniformly from 0 to 1.5 across the 30 patterns. This produces real positive values of \(|C|\), justifying the comparison of \(|C'|\) with \(k(|C|)\). 
Note that the estimator $|C'|=\sqrt{C_I}$ is biased at low correlations, where random fluctuations may produce negative intensity correlations. We therefore restrict the range of \(a\) so that $|C'|$ remains above 0.15.

In the superposed reflections configuration, the reflectivity $\rho$ is controlled by superposing a phase ramp with a six-pixel period on the SLM pattern. The ramp diffracts a controlled fraction of the light in the vertical direction, thereby reducing the effective reflectivity of the SLM in the horizontal direction. The relationship between the amplitude of the phase ramp and the resulting reflectivity is calibrated beforehand, allowing $\rho$ to be controlled with an uncertainty of 0.02. Even though the intrinsic reflectivity of the SLM is $0.65$, the maximum effective reflectivity obtained is $0.44$, owing to losses into the diffraction orders, and a dilution effect due to the increasing size of the diffraction spot after each reflection. 

\section{Summary and Conclusion}

Our starting point was to seek an implementation that would maximize structural nonlinearity by combining full modulation with arbitrarily high orders of interaction with the data. This requirement naturally led to the geometry considered here: a cavity formed by a mirror and an SLM. 
This is an appealing alternative to existing implementations, for its simplicity, maximal nonlinearity, and robustness to thermal drift induced by optical absorption (owing to its globally paraxial geometry). We developed an analytical model from which we obtained closed-form expressions for two established metrics of nonlinearity, the coefficient of determination \(R^2\) and the kernel profile \(k(C)\), directly in terms of the system parameters. In the course of this analysis, we identified a third metric: the average number of interactions with the data, \(\langle n\rangle\), which was found to capture complementary aspects of nonlinearity not captured by \(R^2\), and to be the main parameter governing the kernel profile. This complements other characterization methods of nonlinearity \cite{eliezer2023tunable,han2026optical}, and the question of universality of such architectures \cite{savinson2025universality}, to which it adds a quantitative counterpart, and provides a first-principles design guide for machine learning applications.

Although developed in the context of our particular geometry, our results are more general. The expression of the kernel profile (Eq.~\eqref{gamma}) applies to any system satisfying uniform illumination, normal incidence, and full mixing, whereas the expressions for \(R^2\) (Eq.~\eqref{R2}) and the average number of interactions ($\langle n\rangle=\sum n P_n$) make no assumption on the physical system itself. Importantly, all these quantities are expressed in terms of the order spectrum $P_n$. In a sense, \(P_n\) separates the physics of a particular system from the more general properties of structural nonlinearity. The determination of $P_n$ may be more or less straightforward depending on the implementation; it was found to be deducible from particularly simple arguments in our system, owing to its special geometry, but similar arguments may extend to more complex systems, within and beyond optics.

\section*{Acknowledgment}
The authors thank Yandong Li, Ziao Wang, Laurent Daudet, Baptiste Portail, Thierry Lapointe-Leclerc, Salim Bennani, and Amol Mahurkar for useful discussions. This work was funded by the LION project (216600) of the Swiss National Science Foundation. 

\section*{Data Availability}
The data underpinning this work is available upon request.

\bibliography{aipsamp2.bib}
\clearpage

\section*{APPENDIX A: Derivation of $R^2$}
%\vspace{-5pt}

We adopt a discrete description of the system where light propagates along discrete paths. The output field is  
\begin{equation} \label{descreetdescipt}
\boldsymbol{u}=\sum_{\gamma} \boldsymbol{u}_\gamma, 
\end{equation}
with $\boldsymbol{u}_\gamma$ the contribution of light propagating along a path $\gamma$, which contains a phase modulation determined by the elements of the data visited by $\gamma$. Inserting this into the variance reads
\begin{align}
\operatorname{Var}(\boldsymbol{u}) 
&=
\left\langle
\left|
\sum_\gamma \boldsymbol{u}_\gamma -
\left\langle\sum_\gamma \boldsymbol{u}_\gamma\right\rangle
\right|^2 \right\rangle,
\end{align}
with the brackets denoting averaging over input realizations. For an input being independent and uniformly distributed phases on a $2\pi$ interval, only the paths that do not interact with the input survive the averaging:
\begin{align}
\left\langle\sum_\gamma \boldsymbol{u}_\gamma\right\rangle=\sum_{|\gamma|=0}\boldsymbol{u}_\gamma,
\end{align}
where $|\gamma|$ designates the number of interactions of path $\gamma$ with the input. The variance becomes 
\begin{align}
\operatorname{Var}(\boldsymbol{u}) =
\left\langle \left|\sum_{|\gamma|\geq1}\boldsymbol{u}_\gamma\right|^2\right\rangle .
\end{align}
Expanding the squared modulus and assuming different paths are uncorrelated, the cross terms are negligible and we have 
\begin{align}
\operatorname{Var}(\boldsymbol{u}) = \sum_{|\gamma|\geq1}|\boldsymbol{u}_\gamma|^2.
\end{align}
Note that $\sum_{|\gamma|=n}|\boldsymbol{u}_\gamma|^2$ is the total power of light that interacts $n$ times with the data, which is the definition of our order spectrum $P_n$. Therefore $\operatorname{Var}(\boldsymbol{u}) =1-P_0$, assuming a unit output power such that $\sum_{n=0}^{\infty}P_n=1$. In a similar way the linear regression also subtracts the $|\gamma|=1$ terms, reading 
\begin{align}
\operatorname{Var}(\boldsymbol{u}-W\boldsymbol{x}) =
\sum_{|\gamma|\geq2}|\boldsymbol{u}_\gamma|^2=1-P_0-P_1.
\end{align}
Inserting these into the original expression of $R^2$ we have
\begin{align}
R^2
&=
1-\frac{\operatorname{Var}(\boldsymbol{u}-W\boldsymbol{x})}
{\operatorname{Var}(\boldsymbol{u}) } \\
&=
1-\frac{1-P_0-P_1}{1-P_0} \\
&=
\frac{P_1}{1-P_0}.
\end{align}
QED. Note that this assumes that a linear model does not capture higher order dependencies, which is to say that the different orders in $x$ are orthogonal, or $\langle x^{*n} x^m\rangle_x=\delta[m-n]$. This is a condition on the encoding, rather than the geometry, and is satisfied with our assumed input phase encoding.

\section*{APPENDIX B: Derivation of the kernel profile}
We want to express the correlation of the output field as a function of the correlation of the input data. We use the same path decomposition as above 
\begin{equation}
\boldsymbol{u}=\sum_{\gamma} \boldsymbol{u}_\gamma, 
\end{equation}
with $\boldsymbol{u}_\gamma$ the contribution of path $\gamma$. For the present purpose we need to further decompose $\boldsymbol{u}_\gamma$ as 
\begin{equation}
\boldsymbol{u}_\gamma = \boldsymbol{A}_\gamma \prod_{k\in\gamma} e^{i\phi_{k}},
\end{equation}
with $\boldsymbol{A}_\gamma$ the complex amplitude of the field going through path $\gamma$, and $\phi_{k}$ the phase of the $k$th element of the SLM. Similarly for a different phase pattern $\phi'$ applied to the SLM, the field is 
\begin{equation}
\boldsymbol{u}'_\gamma = \boldsymbol{A}_\gamma \prod_{k\in\gamma} e^{i\phi'_{k}}, 
\end{equation}
The output correlation is proportional to the dot product of the output fields $\boldsymbol{u}$ and $\boldsymbol{u}'$. This yields  
\begin{equation}
C' \propto \boldsymbol{u}^\dagger \boldsymbol{u}'
= \sum_{\gamma,\gamma'} \boldsymbol{u}_\gamma^\dagger \boldsymbol{u}'_{\gamma'}.
\end{equation}
Assuming different paths are uncorrelated, the crossed terms are negligible and we have
\begin{equation}
C' \propto 
\sum_{\gamma}
|\boldsymbol{A}_\gamma|^2
\prod_{k\in\gamma} e^{i\Delta\phi_k},
\end{equation}
with $\Delta\phi_k = \phi'_k - \phi_k$. We now group paths by their number of interactions $n$ with the data and by the sequence of $n$ input elements they visit $s_n$: 
\begin{equation}
C'\propto\sum_n\sum_{s_n}\sum_{\gamma_{s_n}}|
\boldsymbol A_{\gamma_{s_n}}|^2\prod_{k\in s_n} e^{i\Delta\phi_k}.
\end{equation}
The phase product is identical for all paths ${\gamma_{s_n}}$, as they all visit the same sequence of input elements $s_n$. We can therefore factor it out of the sum: 
\begin{equation}
C'\propto
\sum_n\sum_{s_n}\prod_{k\in s_n}e^{i\Delta\phi_{k}}
\sum_{{\gamma_{s_n}}} |\boldsymbol{A}_{{\gamma_{s_n}}}|^2.
\end{equation}
What is left of the sum on ${\gamma_{s_n}}$ is the total power that interacts with a given sequence $s_n$. This can be simplified by assuming full mixing conditions, whereby each reflection redistributes light uniformly across all input data elements, in which case all sequences of a given length $n$ contribute the same power. As there is a total of $M^n$ sequences of length $n$ among $M$ input elements, this sum is simply $P_n/M^n$. This is now only a function of $n$ and can be factored out of the sum on $s_n$:
\begin{comment}
This reveals the quantity 
\begin{equation}
A_{n,s}=\sum_{{\gamma_{s_n}}} |\boldsymbol{A}_\gamma|^2, \end{equation}
which is the total power that interacts with the sequence of pixels $s$. If we assume that $A_{n,s}$ does not depend on $s$, that is, the diffusion is random enough so that all sequences of pixels contribute equally to the output power, we have $A_{n,s}=A_{n}$, which can be factored out of the sum on $s$:
\end{comment}
\begin{equation} \label{someq}
C' \propto\sum_n \frac{P_n}{M^n}
\left( \sum_{s_n} \prod_{k\in s_n} e^{i\Delta\phi_{k}}
\right).
\end{equation}
\clearpage
\noindent The term in parentheses is the sum of all possible (ordered) products of $n$ phase factors. This sum of products is precisely generated by the $n$th power of the sum of these same phase factors
\setlength{\abovedisplayshortskip}{0pt}
\begin{equation}
\sum_{s_n}\prod_{k\in s_n}e^{i\Delta\phi_{k}}
=\left(\sum_{k=1}^{M}e^{i\Delta\phi_k}
\right)^n.
\end{equation}
This reveals the correlation of the input data $C$. Indeed, as $C=\sum_{k=1}^{M}e^{i\Delta\phi_k}/M$, Eq. (\ref{someq}) finally becomes  
\begin{equation} 
C'=\sum_{n=0}^\infty P_n \, C^n, 
\end{equation}
where proportionality becomes equality as $C'(1)=1$. QED.

\vspace{2pt}
\section*{APPENDIX C: Order spectrum of intensity }\label{appD}
\vspace{-10pt}
It is worth noting that, when measuring the output intensity (rather than field), the effective order distribution is no longer $P_n$. Indeed, if the field contains terms of order $n$ and $m$, taking the absolute square generates terms of order $2n$, $2m$, and $n+m$. Measuring intensity therefore redistributes the relative contribution of each order. We can define an intensity order spectrum $P'_k$, analogous to the field order spectrum $P_n$, and ask: can $P'_k$ be expressed in terms of $P_n$? If we gather the values of $P_n$ in a single vector $\boldsymbol{P}$, and similarly with $P'_k$ in a vector $\boldsymbol{P'}$, we find the relationship $\boldsymbol{P}'=\boldsymbol{P}*\boldsymbol{P}$, with $*$ denoting the convolution. This can be shown as follows. Consider the output field expressed as the sum of the contributions from different paths $\gamma$ and number of interactions with the data $n$:
\\
{
\setlength{\abovedisplayskip}{10pt}
\setlength{\belowdisplayskip}{5pt}
\setlength{\abovedisplayshortskip}{3pt}
\setlength{\belowdisplayshortskip}{3pt}
%\vspace{3pt}
\vfill\eject
{\setlength{\abovedisplayskip}{0pt}
\setlength{\abovedisplayshortskip}{0pt}
\begin{equation}\label{sumai}
\boldsymbol{u}=\sum_n\sum_{|\gamma|=n}\boldsymbol{A}_{\gamma},
\end{equation}}
with $P_n$ given by $\sum_{|\gamma|=n} |\boldsymbol{A}_{\gamma}|^2$. Now taking the intensity yields
\begin{equation}
I=|u|^2=\sum_n\sum_m\sum_{|\gamma|=n}\sum_{|\gamma'|=m}
\boldsymbol{A}_{\gamma}\boldsymbol{A}_{\gamma'}^*.
\end{equation}

Treating this sum in an analogous way as Eq.~\eqref{sumai}, we define the intensity order spectrum $P'_k$ as the sum of the absolute square of all terms of order $k$. As $\boldsymbol{A}_{\gamma}\boldsymbol{A}_{\gamma'}^*$ is of order $n+m$, we sum over all terms of order $n+m=k$:
\begin{equation}
P'_k=\sum_{n+m=k}\sum_{|\gamma|=n}\sum_{|\gamma'|=m}
|\boldsymbol{A}_{\gamma}\boldsymbol{A}_{\gamma'}^*|^2.
\end{equation}

The sums over $\gamma$ and $\gamma'$ factor out, revealing our previous definitions of the field order spectra $P_n$ and $P_m$:
\begin{equation}
\begin{split}
P'_k
&=\sum_{n+m=k}\sum_{|\gamma|=n}|\boldsymbol{A}_{\gamma}|^2
\sum_{|\gamma'|=m}|\boldsymbol{A}_{\gamma'}|^2,\\
&=\sum_{n+m=k}P_n P_m.
\end{split}
\end{equation}
which is simply the convolution of $P_n$ with itself (the self-convolution) evaluated at $k$, hence $\boldsymbol{P}'=\boldsymbol{P}*\boldsymbol{P}$. This result can be seen as the classical product--convolution duality of Fourier analysis, in disguise.
}

\vspace{-10pt}
\section*{APPENDIX D: More kernel profiles }
\vspace{-10pt}
Here we present additional experimental measurements of the kernel profile for different values of $\rho$ and $N$ in both configurations together with their predicted profiles. 
\vspace{10pt}

\addtolength{\textheight}{3\baselineskip}
\onecolumngrid\refstepcounter{figure}\renewcommand{\thefigure}{A1}
\noindent\includegraphics[width=\textwidth]{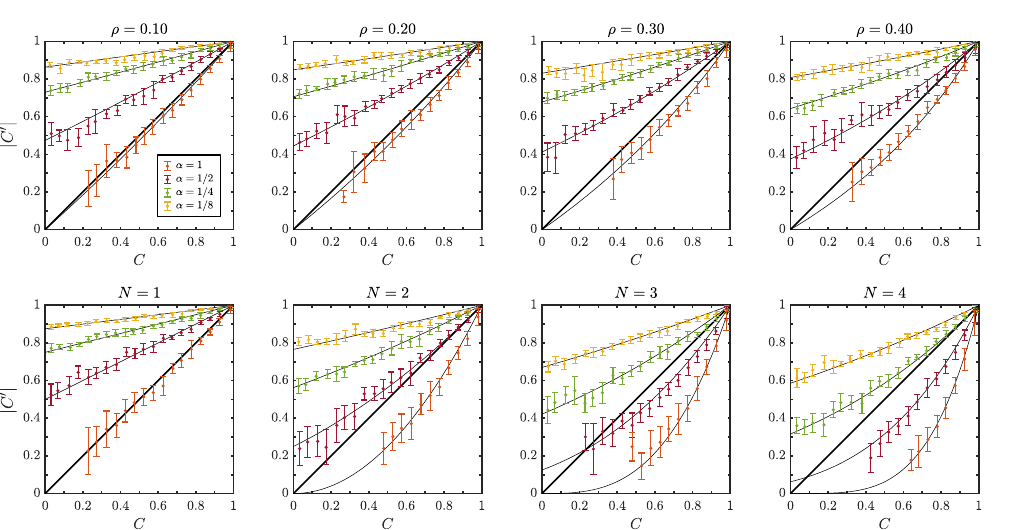}
\noindent\textbf{FIG.~\thefigure.} Measured kernel profiles for superposed reflections (top row) and isolated reflections (bottom row), for different values of $\rho$ and $N$, respectively. In each case, the profile is measured for different values of $\alpha$ (the fraction of modulated area). The thin line shows the profile predicted by Eq.~\eqref{kernelsuperp} and Eq.~\eqref{kernelisol}. 
\label{}
\twocolumngrid

\end{document}